\documentclass{sig-alternate}
\usepackage[ruled,vlined]{algorithm2e}
\usepackage{graphicx}
\usepackage{url}
\usepackage{paralist}
\usepackage{tabularx}
\usepackage{balance}
\usepackage{multirow}
\usepackage{multicol}
\usepackage{pbox}
\usepackage{mathtools}
\usepackage[TABBOTCAP]{subfigure}
\usepackage{algorithmic}
\usepackage{graphicx}
\usepackage{paralist}
\usepackage{tabularx}
\usepackage{url}
\usepackage{balance}
\usepackage{multirow}
\usepackage{multicol}
\usepackage{amsmath}
\usepackage{setspace}
\usepackage{hyperref}
\usepackage{verbatim}
\usepackage{float}
\usepackage[normalem]{ulem}
\usepackage{xspace}
\usepackage[usenames]{color}
\usepackage{colortbl}
\usepackage{amssymb}
\usepackage{ifthen}
\usepackage{pifont}
\usepackage{tikz}
\newcommand*\circled[1]{\tikz[baseline=(char.base)]{
            \node[shape=circle,draw,inner sep=0.5pt] (char) {#1};}}

\include{macro}

\begin{document}
\toappear{}

\pagenumbering{gobble}

\title{Auto-completing Bug Reports for Android Applications}

\numberofauthors{1} 
\author{
\alignauthor
Kevin Moran, Mario Linares-V\'asquez, Carlos Bernal-C\'ardenas, Denys Poshyvanyk\\
\affaddr{College of William \& Mary}\\
\affaddr{Department of Computer Science}\\
\affaddr{Williamsburg, VA 23187-8795, USA}\\
\email{\{kpmoran, mlinarev, cebernal, denys\}@cs.wm.edu}
}
\maketitle
\begin{abstract}
The modern software development landscape has seen a shift in focus toward mobile applications as tablets and smartphones near ubiquitous adoption. Due to this trend, the complexity of these ``apps" has been increasing, making development and maintenance challenging. Additionally, current bug tracking systems are not able to effectively support construction of reports with actionable information that directly lead to a bug's resolution. To address the need for an improved reporting system, we introduce a novel solution, called FUSION, that helps users auto-complete reproduction steps in bug reports for mobile apps. FUSION links user-provided information to program artifacts extracted through static and dynamic analysis performed before testing or release. The approach that FUSION employs is generalizable to other current mobile software platforms, and constitutes a new method by which off-device bug reporting can be conducted for mobile software projects.  In a study involving 28 participants we applied FUSION to support the maintenance tasks of reporting and reproducing defects from 15 real-world bugs found in 14 open source Android apps while qualitatively and qualitatively measuring the user experience of the system. Our results demonstrate that FUSION both effectively facilitates reporting and allows for more reliable reproduction of bugs from reports compared to traditional issue tracking systems by presenting more detailed contextual app information.
\end{abstract}
\vspace{-2mm}

\vspace{-2mm}
\vspace{-2mm}
\\
\\

\section{Introduction}
\label{sec:intro}
	Smartphones and mobile computing have skyrocketed in popularity in recent years, and adoption has reached near-ubiquitous levels with over 2.7 billion active smartphone users in 2014 \cite{24MobilityReport}.  An increased demand for high-quality, robust mobile applications is being driven by a growing user base that performs an increasing number of computing tasks on ``smart'' devices. Due to this demand, the complexity of mobile applications has been increasing, making development and maintenance challenging. The intense competition present in mobile application marketplaces like Google Play and the Apple App Store, means that if an app is not performing as expected,  due to bugs or lack of desired features, 48\% of users are less likely to use the app again and will abandon it for another one with similar functionality \cite{app-abandonment}. 
	
	Software maintenance activities are known to be generally expensive and challenging \cite{25Tassey:NIST}.  One of the most important maintenance tasks is bug report resolution.  However, current bug tracking systems such as Bugzilla \cite{bugzilla}, Mantis \cite{mantis}, the Google Code Issue Tracker \cite{google-code}, the GitHub Issue Tracker \cite{github-it}, and commercial solutions such as JIRA \cite{jira} rely mostly on unstructured natural language bug descriptions.  These descriptions can be augmented with files uploaded by the reporters (e.g., screenshots). As an important component of bug reports, reproduction steps are expected to be reported in a structured and descriptive way, but the quality of description mostly depends on the reporter's experience and attitude towards providing enough information. Therefore, the reporting process can be cumbersome, and the additional effort means that many users are unlikely to enhance their reports with extra information \cite{11Bettenburg:MSR08,31Davies:ESEM2014,32Bettenburg:ICSM08, 34Aranda:ICSE09}. 
	
	 A past survey of open source developers conducted by Koru et al. has shown that only  $\approx$ 50\% of developers believe bug reports are always complete \cite{33Koru:IEEE2004}.  Previous studies have also shown that the information most useful to developers is often the most difficult for reporters to provide and that the lack of this information is a major reason behind non-reproducible bug reports \cite{4Joorabchi:MSR14, 3Bettenburg:FSE08}.   Difficulty providing such information, especially reproduction steps, is compounded in the context of mobile applications due to their complex event-driven and GUI-based nature.  
	 Furthermore, many bug reports are created from textual descriptions of problems in user reviews.  According to a recent study by Chen et al. \cite{Chen:icse2014}, only a reduced set of user reviews can be considered useful and/or informative. Also, unlike issue reports, reviews do not refer to app implementation details.
	
	The above issues point to a more prominent problem for bug tracking systems in general: the \textit{lexical gap} that normally exists between bug reporters (e.g., testers, beta users) and developers.  Reporters typically only have functional knowledge of an app, even if they have development experience themselves, whereas the developers working on an app tend to have intimate code level knowledge.  In fact, a recent study conducted by Huo et al. corroborates the existence of this knowledge gap as they found there is a difference between the way experts and non-experts write bug reports as measured by textual similarity metrics \cite{35Huo:ICSME14}.  When a developer reads and attempts to comprehend (or reproduce) a bug report, she has to bridge this gap, reasoning about the code level problems from the high-level functional description in the bug report.  If the lexical gap is too wide the developer may not be able to reproduce and/or subsequently resolve the bug report. 	

	To address this fundamental problem of making bug reports more useful (and reproducible) for developers, we introduce a novel approach, which we call FUSION, that relies on a novel \textit{Analyze $\rightarrow$ Generate} paradigm to enable the auto-completion of Android bug reports in order to provide more actionable information to developers.  In the context of this work, we define auto-completion as suggesting relevant actions, screen-shots, and images of specific GUI-components to the user in order to facilitate reporting the steps for reproducing a bug.  FUSION first uses fully automated static and dynamic analysis techniques to gather screen-shots and other relevant information about an app before it is released for testing. Reporters then interact with the web-based report generator using the auto-completion features in order to provide the bug reproduction steps.  By linking the information provided by the user with features extracted through static and dynamic analyses, FUSION presents an augmented bug report to the developer that contains immediately actionable information with well-defined steps to reproduce a bug. 
	
	We evaluate FUSION in a study comparing bug reports submitted using our system to bug reports produced using Google Code Issue Tracker, involving 28 participants reporting bugs for 15 real-world failures stemming from 14 open source Android apps. 
	
	Our paper makes the following noteworthy contributions:
	 \begin{enumerate}
\item We design and implement a novel approach for auto-completing and augmenting Android bug reports, called FUSION, which leverages static and dynamic analyses, and provides actionable information to developers. The tool facilitates the reporting, reproduction and subsequent resolution of Android bugs. The program analysis techniques of the apps can be run on \textit{both} physical devices and emulators;
\item We design and carry out a comprehensive user study to evaluate the \textit{user experience} of our approach and the \textit{quality} of bug reports generated using FUSION compared to the Google Code Issue Tracker. The results of this study  demonstrate that FUSION enables developers to submit bug reports that are more likely to be reproducible compared to reports written entirely in natural language;  
\item We make FUSION and all the data from the experiments available for researchers \cite{appendix} in hope that this work spurs new research related to improving the quality of bug reports and bug reporting systems.  
\end{enumerate}

\section{State of Research and Practice}
\label{sec:related works}
	 Bug and error reporting has been an active area of research in the software engineering community.  However, little work has been conducted to improve the lack of structure in the reporting mechanism for entering reproduction steps, and adding corresponding support in bug tracking systems. Therefore, in this section, we briefly survey the features of current bug reporting systems and the studies that motivated this work.  Then we differentiate our work from approaches for reproducing in-field failures and explain how our work compliments existing research on bug reporting.

\subsection{Existing Bug Reporting Systems}
	Most current issue tracking systems rely upon unstructured natural language descriptions in their reports.  However, some systems do offer more functionality.  For instance, the Google Code Issue Tracker (GCIT) \cite{google-code} offers a semi-structured area where reporters can enter reproduction steps and expected input/output in natural language form (i.e., the online form asks: ``What steps will reproduce the problem?").  Nearly all current issue trackers offer structured fields to enter information such as tags, severity level, assignee, fix time, and product/program specifications.  Some web-based bug reporting systems (e.g. Bugzilla \cite{bugzilla}, Jira \cite{jira}, Mantis \cite{mantis}, UserSnap \cite{usersnap}, BugDigger  \cite{22BugDigger}) facilitate reporters including screenshots.  However, current bug tracking systems do not integrate online suggestion of relevant reproduction steps with screenshots as FUSION does.

\subsection{Bug Reporting Studies}
	The problem facing many current bug reporting systems is that typical natural language reports capture a coarse grained level of detail that makes developer reasoning about defects difficult.
		This highlights the underlying \textit{task} that bug reporting systems must accomplish: \textit{bridging the lexical knowledge gap between typical reporters of a bug and the developers that must resolve the bugs.}  	Previous studies on bug report quality and developer information needs highlight several factors that can impact the quality of bug reports \cite{15Breu:CSCW10, 4Joorabchi:MSR14, 3Bettenburg:FSE08}:

\begin{itemize}
	\item Other than ``Interbug dependencies'' (i.e., a situation where a bug was fixed in a previous patch), \textit{insufficient information} is one of the leading causes of non-reproducible bug reports \cite{4Joorabchi:MSR14};
	\item Developers consider (i)\textit{steps to reproduce}, (ii)\textit{stack traces}, and (iii)\textit{test cases/scenarios} as the most helpful sources of information in a bug report \cite{3Bettenburg:FSE08};
	\item Information needs are greatest early in a bug's life cycle, therefore, a way to easily add the above features is important during bug report creation \cite{15Breu:CSCW10}.
\end{itemize}

Using these issues as motivation, we developed FUSION with two major goals in mind: (i) \textit{provide bug reports to developers with immediately actionable knowledge (reliable reproduction steps)} and (ii) \textit{facilitate reporting by providing this information through an auto-completion mechanism.}

It is worth noting that one previous study conducted by Bhattacharya et$.$ al$.$ \cite{5Bhattacharya:CSMR13} concluded that most Android bug reports for open source apps are of high-quality, however in their study only $\approx$ 46\% of bug report contained steps to reproduce, and even fewer ($\approx$ 20\%) contained additional information (e.g. bug-triggering input or even an app version).  Therefore, there is clearly room for improvement in terms of the type of information that is contained within open source Android bug reports.

\subsection{In-Field Failure Reproduction}

	A body of work known as in-field failure reproduction \cite{27Bell:ICSE13, 26Jin:ISSTA13, 8Zhou:ICSE12, 29Clause:ICSE07,18Jin:ICSE12,41Artzi:ECOOP2008,50Kifetew:ICST2014,43Cao:ASE14} shares similar goals with our approach. These techniques collect run-time information (e.g., execution traces) from instrumented programs that provide developers with a better understanding of the causes of an in-field failure, which will subsequently help expedite the fixing of those failures.  However, there are several key differences that set our work apart and illustrate how FUSION improves upon the state of research.   

	\textit{First}, techniques regarding in-field failure reproduction rely on potentially expensive program instrumentation, which requires developers to modify code and introduce overhead.  FUSION is completely automatic; our static and dynamic analysis techniques only need to be applied once for the version of the program that is released for testing.  Furthermore, the analysis process can be done without the need for instrumentation of programs in the field.  \textit{Second}, current in-field failure reproduction techniques require an oracle to signify when a failure has occurred (e.g., a crash).  FUSION is not an approach for crash or failure detection; it is designed to support testers during the bug reporting process. \textit{Third}, these techniques have not been applied to mobile apps and would most likely need to be optimized further to be applicable for the corresponding resource-constrained enviornment.  

\subsection{Bug and Error Reporting Research}
	A subset of prior work has focused on bug and crash triage \cite{7Shokripour:MSR13, 10Naguib:MSR13,40Jeong:FSE2009,44Kim:TOSE2013,45Park:AAAI2011,46Kim:DSN2011,51Haihao:ICST2011,34Aranda:ICSE09,Linares-Vasquez:ICSM2012,Menzies:ICSM2008}. 
	The techniques associated with this topic typically employ different program analysis and machine learning or natural language processing techniques to match bug reports with appropriate developers.  Our proposed research compliments developer recommendation frameworks, as FUSION can provide these frameworks with more detailed ``knowledge'' than current state of practice bug reporting systems.  

	A significant amount of research has been conducted concerning the summarization \cite{1Mani:FSE12,11Bettenburg:MSR08,20Rastkar:ICSE10,33Koru:IEEE2004,53Weiss:MSR2007,Czarnecki:ICSM2012}, fault localization \cite{8Zhou:ICSE12,13Wang:ICPC14,37Rahman:FSE2011,38Baudry:ICSE2006,39Vidacs:CSMR14,42Wu:ISSTA2014,47Marsi:STVR2010,49Ayewah:IS2008,52Cleve:ICSE2005,54Dallmeier:LNCS2005}, classification and detection of duplicate bug reports \cite{4Joorabchi:MSR14, 14Nguyen:ASE12, 17Wang:ICSE08, 19Guo:ICSE10,  21Zhou:CIKM12, 36Gu:ICSE10,48Podgurski:ICSE2003}.
	Again, the work presented in this paper compliments these categories of research as bug reports created with FUSION can provide more detailed information, easily linking the bug back to source code, allowing for better localization, summarization and, potentially, duplicate detection.  It is worth noting that work by Bettenburg et$.$ al$.$ on extracting structural information from bug reports is also related; however, we aim at helping auto-complete the structured reproduction steps at the time of report creation, rather than extracting it after the fact \cite{11Bettenburg:MSR08}.

\section{The FUSION Approach}
\label{sec:approach}
\begin{figure*}[tb]
\centering
\includegraphics[width=0.99\linewidth]{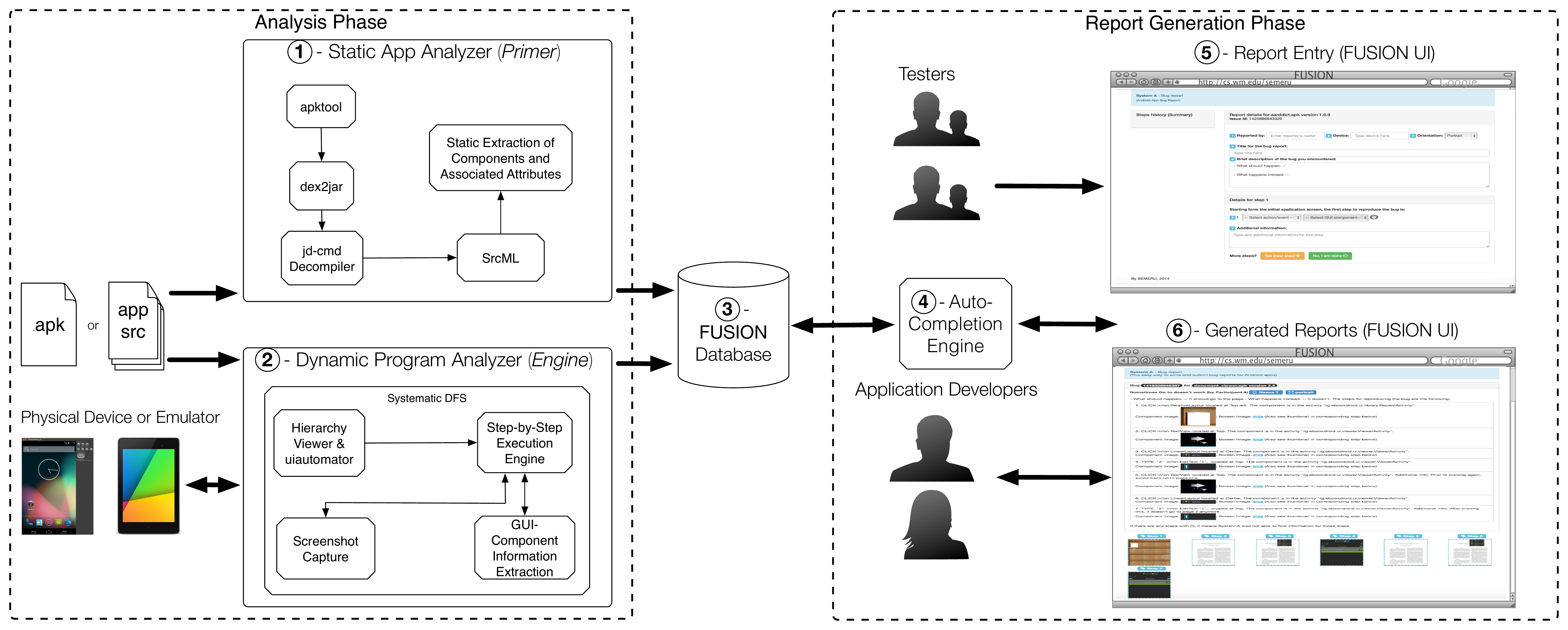}
\vspace{-0.3cm}
\caption{Overview of FUSION Workflow}
\label{Design}
\vspace{-0.3cm}
\end{figure*}

    FUSION's \textit{Analyze} $\rightarrow$ \textit{Generate} workflow corresponds to two major phases. In the \textit{Analysis Phase} FUSION collects information related to the GUI components and event flow of an app through a combination of static and dynamic analysis.  Then in the \textit{Report Generation Phase} FUSION takes advantage of the GUI centric nature of mobile apps to both auto-complete the steps to reproduce the bug and augment each step with contextual application information.  The overall design of FUSION can be seen in Figure \ref{Design}.  We encourage readers to view videos of our tool in use, complete with commentary, available in our replication package outlined in Section \ref{sec:rep-pack} and online at \cite{appendix}.
    
\subsection{Analysis Phase}
 
    The \textit{Analysis Phase} collects all of the information required for the \textit{Report Generation Phase} operation. This first phase has two major components: 1) static analysis \emph{(Primer)}, and 2) dynamic program analysis \emph{(Engine)} of a target app. 
    The \textit{Analysis Phase} must be performed before each version of an app is released for testing or before it is published to end users.  Both components of the \emph{Analysis Phase} store their extracted data in the FUSION database (Fig. \ref{Design} - \circled{3}).
    
\subsubsection{Static Analysis (Primer)}
    
    The goal of the \emph{Primer} (Fig. \ref{Design} - \circled{1})
   is to extract all of the GUI components and associated information from the app source code.   For each GUI component, the \emph{Primer} extracts: (i) possible actions on that component, (ii) type of the component (e.g., Button, Spinner), (iii) activities the component is contained within, and (iv) class files where the component is instantiated.  
    Thus, this phase gives us a universe of possible components within the domain of the application, and establishes traceability links connecting GUI components that reporters operate upon to code specific information such as the class or activity they are located within.  
    
    The \emph{Primer} is comprised of several steps to extract the information outlined above. First it uses the  {\tt dex2jar}\cite{dex2jar} and  {\tt jd-cmd} \cite{jd-cmd}  tools for decompilation; then, it converts the source files to an XML-based representation using {\tt srcML} \cite{srcml}. We also use {\tt apktool} \cite{apktool} to extract the resource files from the app's APK.  The {\tt id}s and types of GUI components were extracted from the xml files located in the app's resource folders (\textit{i.e.},  {\tt /res/layout} and  {\tt /res/menu} of the decompiled application or src).  Using the {\tt srcML} representation of the source code, we are able to parse and link the GUI-component information to extracted app source files.  
    
\subsubsection{Dynamic Analysis (Engine)}

The \emph{Engine} (Fig. \ref{Design} - \circled{2}) is used to glean dynamic contextual information, such as the location of the GUI component on the screen, and enhance the database with both run-time GUI and application event-flow information. The goal of the \emph{Engine} is to explore an app in a systematic manner, ripping and extracting run-time information related to the GUI components during execution including: (i) the text associated with different GUI components (e.g., the ``Send'' text on a button to send an email message), (ii) whether the GUI component triggers a transition to a different activity, (iii) the action performed on the GUI component during systematic execution, (iv) full screen-shots before and after each action is performed, (v) the location of the GUI component object on the test device's screen, (vi) the current activity and window of each step, (vii) screen-shots of the specific GUI component, and (viii) the object index of the GUI component (to allow for differentiation between different instantiations of the same GUI component on one screen).

    The \emph{Engine} performs this systematic exploration of the app using the {\tt UIAutomator} \cite{uiautomator} framework included in the Android SDK.  This systematic execution of the app is similar to existing approaches in GUI ripping \cite{Takala:ICST2011,Ravindranath:Mobisys2014,Amalfitano:ASE2012,Azim:OOPSLA2013,Machiry:FSE2013,Choi:OOPSLA2013,Nguyen:TOSE2014}. Using the {\tt UIAutomator} framework allows us to capture cases that are not captured in previous tools such as pop-up menus that exist within menus, internal windows, and the onscreen keyboard.  To effectively explore the application we implemented our own version of a systematic depth-first search (DFS) algorithm for application traversal that performs click events on all the clickable components in the GUI hierarchy reachable using the DFS heuristic.
      
    During the ripping, before each step is executed on the GUI, the \emph{Engine} calls {\tt UIAutomator} 
     subroutines to extract the contextual information outlined above regarding each currently displayed GUI component.  We then execute the action associated with each GUI component in a depth-first manner on the current screen. Our current implementation of DFS only handles the click/tap action; however, as this is the most common action, it is still able to explore a significant amount of an application's functionality.  
     
     In the DFS algorithm, if a link is clicked that would normally transition to a screen in an external activity (e.g., clicking a web link that would launch the Chrome web browser app), we execute a \textit{back} command in order to stay within the current app.  If the DFS exploration exits the app to the home screen of the device/emulator for any reason, we simply re-launch the app and  continue the GUI traversal.  During the DFS exploration, the \emph{Engine} captures every activity transition that occurs after each action is performed (e.g., whether or not a new activity is started/resumed after an action to launch a menu).  This allows FUSION to build a model of the app execution that we will later use to help track a reporter's relative position in the app when they are using the system to record the steps to reproduce a bug.  
\subsection{Report Generation Phase}  
         
         We had two major goals when designing the \textit{Report Generation Phase} component of FUSION:
    
\begin{enumerate}
  \item Allow for traditional natural language input in order to give a high-level overview of a bug.
  \item Auto-complete the reproduction steps of a bug through suggestions derived by tracking the position of the reporter's step entry in the app event-flow.
\end{enumerate}
       
                During the \emph{Report Generation Phase}, FUSION aids the reporter in constructing the steps needed to recreate a bug by making suggestions based upon the ``potential" GUI state reached by the declared steps.  This means for each step $s$, FUSION infers --- online --- the GUI state $GUI_s$ in which the target app should be by taking into account the history of steps.   For each step, FUSION verifies that the suggestion made to the reporter is correct by presenting the reporter with contextually relevant screen-shots, where the reporter selects the screen-shot corresponding to the current action the reporter wants to describe.

\subsubsection{Report Generator User Interface}

\begin{figure}[t]
\centering
\includegraphics[width=\linewidth]{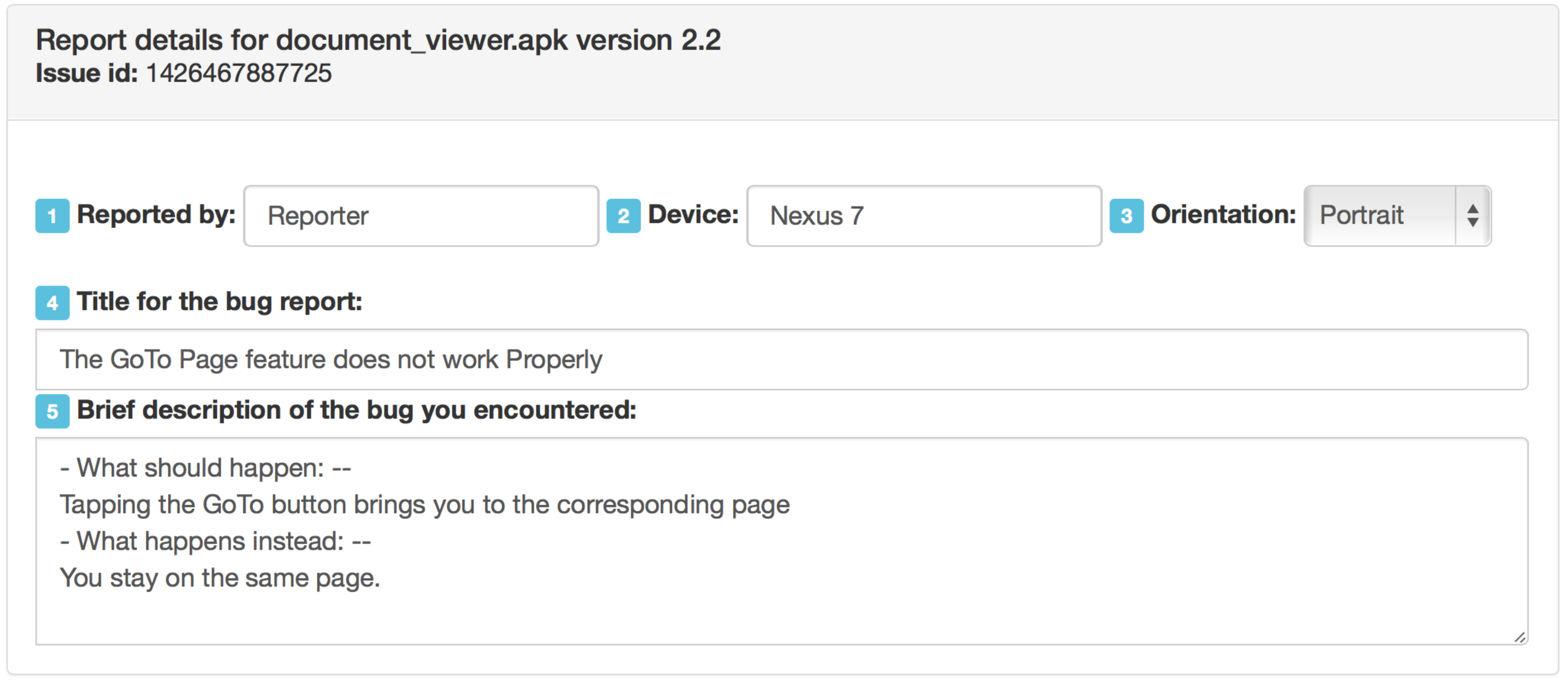}
\vspace{-0.5cm}
\caption{FUSION Reporter Interface}
\label{front_end}
\end{figure}

    After first selecting the app to report an issue for, a reporter interacts with FUSION by filling in some identifying information (i.e., name, device, title) and a brief textual description of the bug in question in the top half of the UI.  Next, the reporter inputs the steps to reproduce the bug using the auto-completion boxes in a step-wise manner, starting from the initial screen of a cold app launch\footnote{Cold-start means the first step is executed on the first window and screen displayed directly after the app is launched.}, and proceeds until the list of steps to reproduce the bug is exhausted.   Let us consider a running example where the user is filling out a report for the Document Viewer bug in Table \ref{tab:bug-reports}.  According to the various fields in Figure \ref{front_end}, the reporter would first fill in their (i) \textit{name} (Field 1), (ii) \textit{device} (Field 2), (iii) \textit{screen orientation} (Field 3), (iv) a \textit{bug report title} (Field 4), and (v) a \textit{brief description of the bug} (Field 5). 
    		
\subsubsection{Auto-Completing Bug Reproduction Steps}
    
     To facilitate the reporter in entering reproduction steps, we model each step in the reproduction process as an {\tt \{action, component\}} tuple corresponding to the action the reporter wants to describe at each step, (e.g., tap, long-tap, swipe, type) and the component in the app GUI with which they interacted (e.g.,``Name" textview, ``OK" button, ``Days" spinner).  Since reporters are generally aware of the actions and GUI elements they interact with, it follows that this is an intuitive manner for them to construct reproduction steps.  FUSION allocates auto-completion suggestions to drop down lists based on a decision tree taking into account a reporter's position in the app execution beginning from a cold-start of the app.
     
\begin{figure}[h]
\centering
\includegraphics[width=\linewidth]{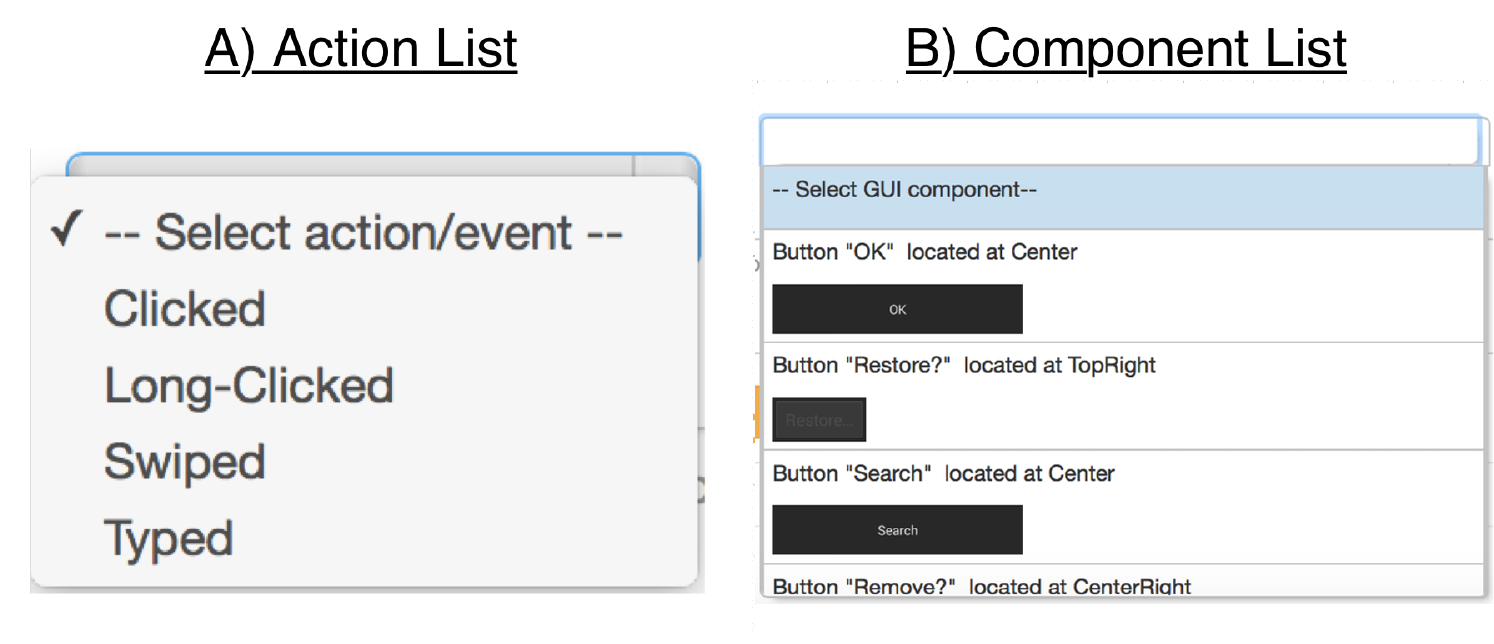}
\vspace{-0.7cm}
\caption{Auto-Complete Dropdown Menus}
\label{autocomplete_box}
\end{figure}

     The first drop down list (see Figure \ref{autocomplete_box}-A) corresponds to the possible actions a user can perform at a given point in app execution.  In our example with the Document Viewer bug, let's say the reporter selects \textit{click} as the first action in the sequence of steps as shown in Figure \ref{autocomplete_box}-A. The possible actions considered in FUSION  are \textit{click(tap), long-click(long-touch), type}, and \textit{swipe}.  The \textit{type} action corresponds to a user entering information from the device keyboard.  When the reporter selects the \textit{type} option, we also present them with a text box to collect the information she typed in the Android app.

	The second dropdown list (see Figure \ref{autocomplete_box}-B) corresponds to the component associated with the action in the step. FUSION presents the following information, which can also be seen in Figure \ref{autocomplete_box}: (i) \textit{Component Type}: this is the type of component that is being operated upon, e.g., button, spinner, checkbox,  (ii) \textit{Component Text}: the text associated with or located on the component, (iii) \textit{Relative Location}: the relative location of the component on the screen according to the parameters in Figure \ref{example_screenshot}, and (iv) \textit{Component Image}: an in-situ (i.e., embedded in the dropdown list) image of the instance of the component.  The relative location is displayed here to make it easier for reporters to reason about the on-screen location, rather than reasoning about pixel values.  In our running example, FUSION will populate the component dropdown list with all of the clickable components in the Main Activity since this is the first step and the selected action was \textit{click}.  The user would then select the component they acted upon, in this case, the first option in the list: the ``OK" button located at the center of the screen (see Figure \ref{autocomplete_box}-B).
    
    One potential issue with component selection from the auto-complete drop-down list is that there may be duplicate components on the same screen in an app.  FUSION solves this problem in two ways. \textit{First}, it differentiates each duplicate component in the list through specifying text ``Option \#''.  \textit{Second} FUSION attempts to confirm the component entered by the reporter at each step by fetching screen-shots from the FUSION database representing the entire device screen.  Each of these screen-shots highlights the representative GUI component as shown in Fig. \ref{example_screenshot}. To complete the step entry, the reporter simply selects the screen-shot corresponding to both the app state and the GUI component acted upon.  In our running example, the reporter would select the full augmented screenshot corresponding to the component they selected from the dropdown list. In our case, an illustrative portion of the screenshot for the ``OK" button is shown in Figure \ref{example_screenshot}.
    
     After the reporter makes selections from the drop-down lists, they have an opportunity to enter additional information for each step (e.g., a button had an unexpected behavior) in a natural language text entry field.  For instance in our running example, the reporter might indicate that after pressing the ``OK" button the pop-up window took longer than expected to disappear.

   \subsubsection{Report Generator Auto-Completion Engine}
 
 The \emph{Auto-Completion Engine} of the web-based report generator (Figure \ref{Design}-\circled{4}) uses the information collected up-front during the \textit{Analysis Phase}. When FUSION suggests completions for the drop-down menus, it queries the database for the corresponding state of the app event flow and suggests information based on the past steps that the reporter has entered.  Since we always assume a ``cold'' application start, the \emph{Auto-Completion Engine} starts the reproduction steps entry process from the app's main Activity.  We then track the reporter's progress through the app using predictive measures based on past steps.  
 
\begin{figure}[t]
\centering
\vspace{+.02cm}
\includegraphics[width=\linewidth]{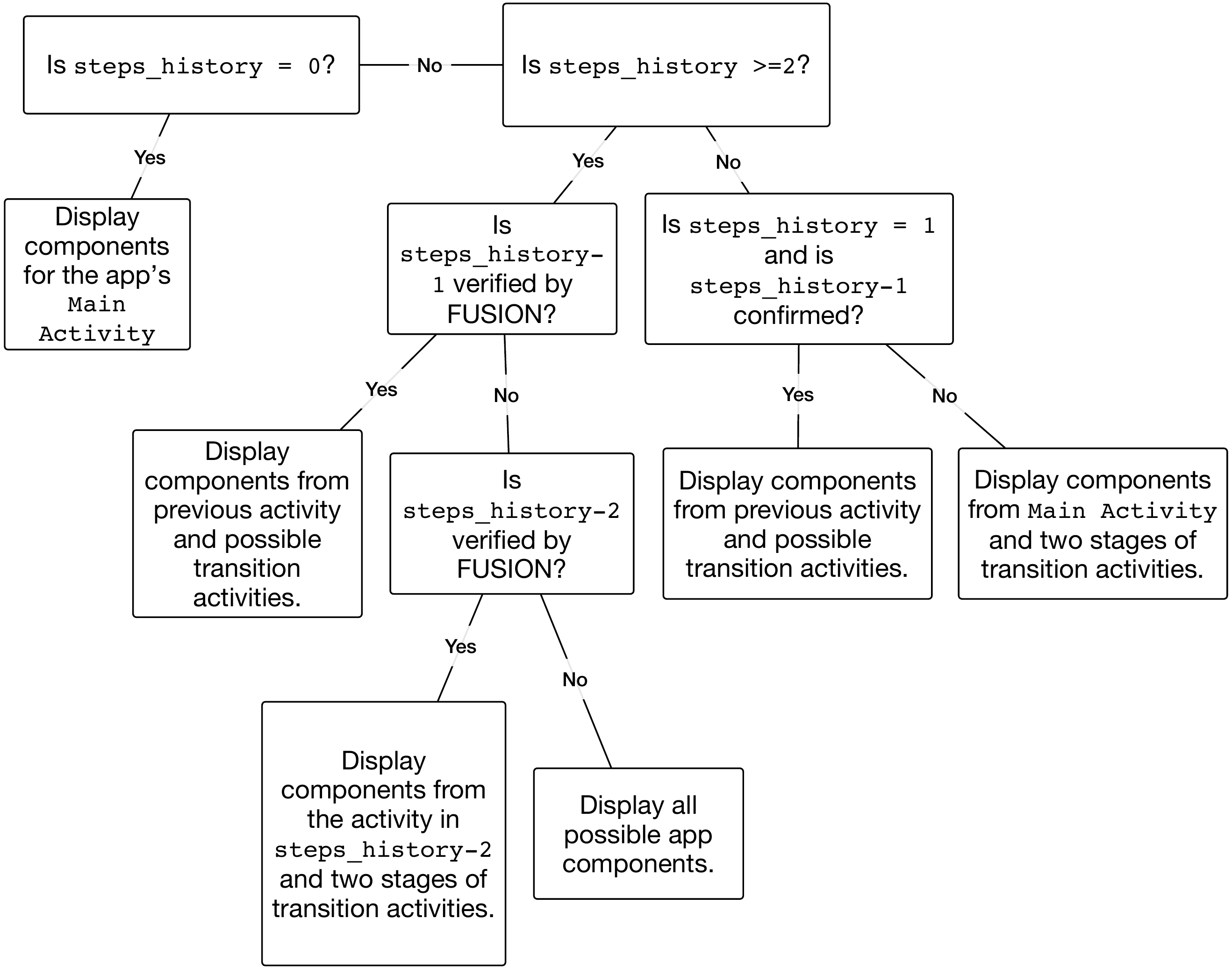}
\vspace{-0.5cm}
\caption{Decision Tree Utilized by Auto-Completion Engine}
\label{decision_tree}
\end{figure}

 	The \emph{Auto-Completion Engine} operates on application steps using several different pieces of information as input.  It models the reporter's reproduction steps as an ordered stream of steps $S$ where each individual step $s_i$ may be either empty or full.  Each step can be modeled as a five-tuple consisting of \textit{\{step\_num, action, comp\_name, activity, history\}}.  The \textit{action} is the gesture provided by the reporter in the first drop-down menu.  The \textit{component\_name} is the individual component name as reported by the {\tt UIautomator} interface during the Engine phase.  The \textit{activity} is the Android screen the component is found on.  The \textit{history} is the history of steps preceding the current step.  The auto-completion engine predicts the suggestion information using the decision tree logic which can be seen in Figure \ref{decision_tree}.  
	
	FUSION presents components to the reporter at the granularity of activities or application screens.  To summarize the suggestion process, FUSION looks back through the history of the past few steps and looks for possible transitions from the previous steps to future steps depending on the components interacted with.  If FUSION is unable to capture the last few steps from the reporter due to the incomplete application execution model mentioned earlier, then FUSION presents the possibilities from all known screens of the application. In our running example, let's consider the reporter moving on to report the second reproduction step.  In this case, FUSION would query the history to find the previous activity the ``OK" button was located within, and then present component suggestions from that activity, in the case that the user stayed in the same activity, and the components from possible transition activities, in the case the user transitioned to a different activity.

\subsubsection{Handling FUSION's Application Model Gaps}
    
\begin{figure}[t]
\centering
\includegraphics[width=\linewidth]{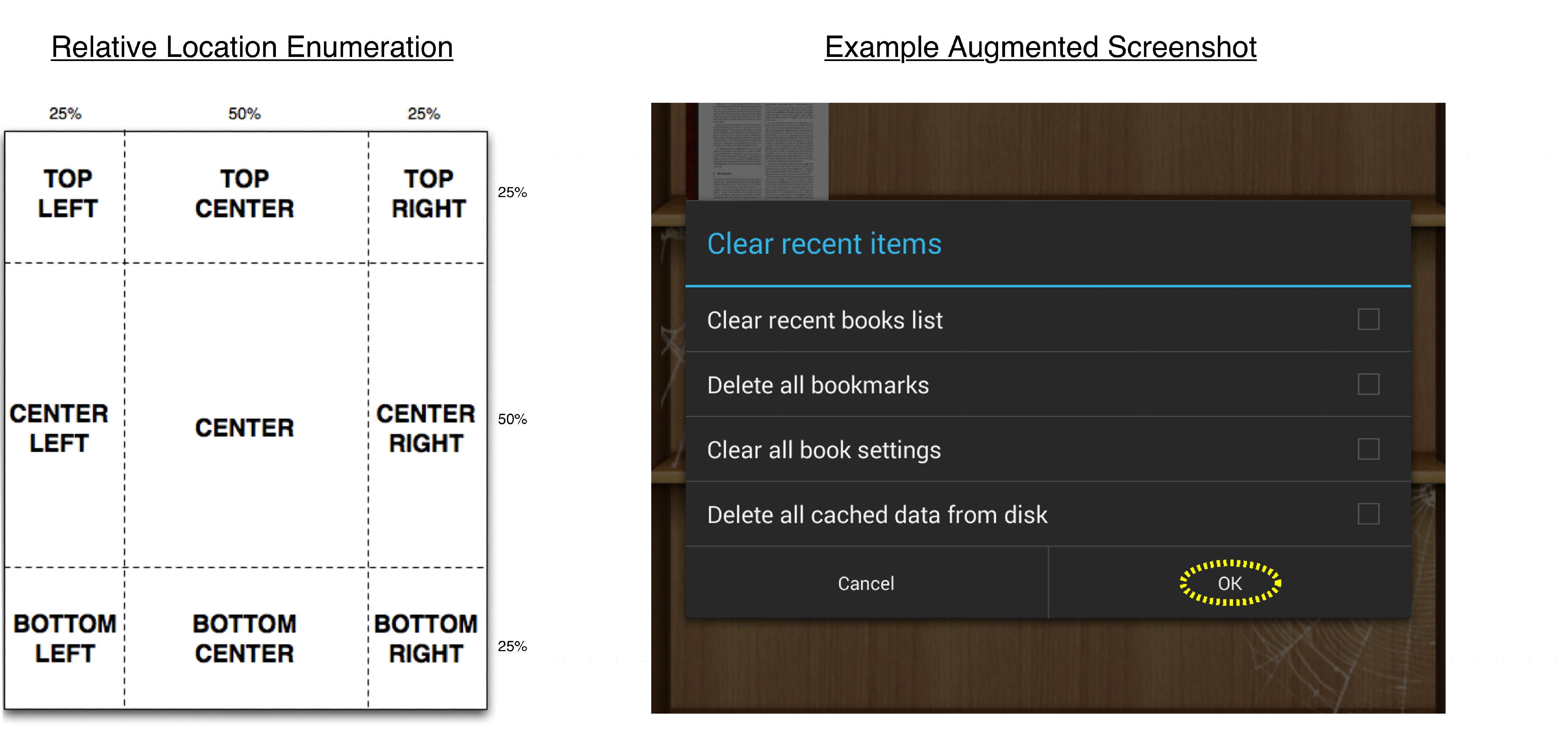}
\vspace{-0.8cm}
\caption{Relative Location Enumeration and Example Augmented Screenshot}
\label{example_screenshot}
\end{figure}

    Because DFS-based exploration is not exhaustive \cite{Nguyen:TOSE2014}, there may be gaps in FUSION's database of possible app screens (e.g., a dynamically generated component that triggers an activity transition was not acted upon). Due to this, a reporter may not find the appropriate suggestion in the drop-down list.  To handle these cases gracefully, we allow the reporter to select a special option when they cannot find the component they interacted with in the auto-complete drop-down list. In our running example, let's say the reporter wishes to indicate that they clicked the button labeled ``Open Document," but the option is not available in the auto-complete component drop-down list.  In this case the reporter would select the ``Not in this list...'' option and manually fill in  (i) the type of the component (to limit confusion, we present this option as a drop-down box auto-completed with only the GUI-component types that exist in the application, as extracted by the \emph{Primer}, in our case the user would choose ``Button"), (ii) any text associated with the GUI-component (in this case ``Open Document") and (iii) the relative location of the GUI-component as denoted in Figure \ref{example_screenshot} (in this case ``Top Center").
    
\subsubsection{Report Structure}

\begin{figure}[t]
\centering
\includegraphics[width=0.95\linewidth]{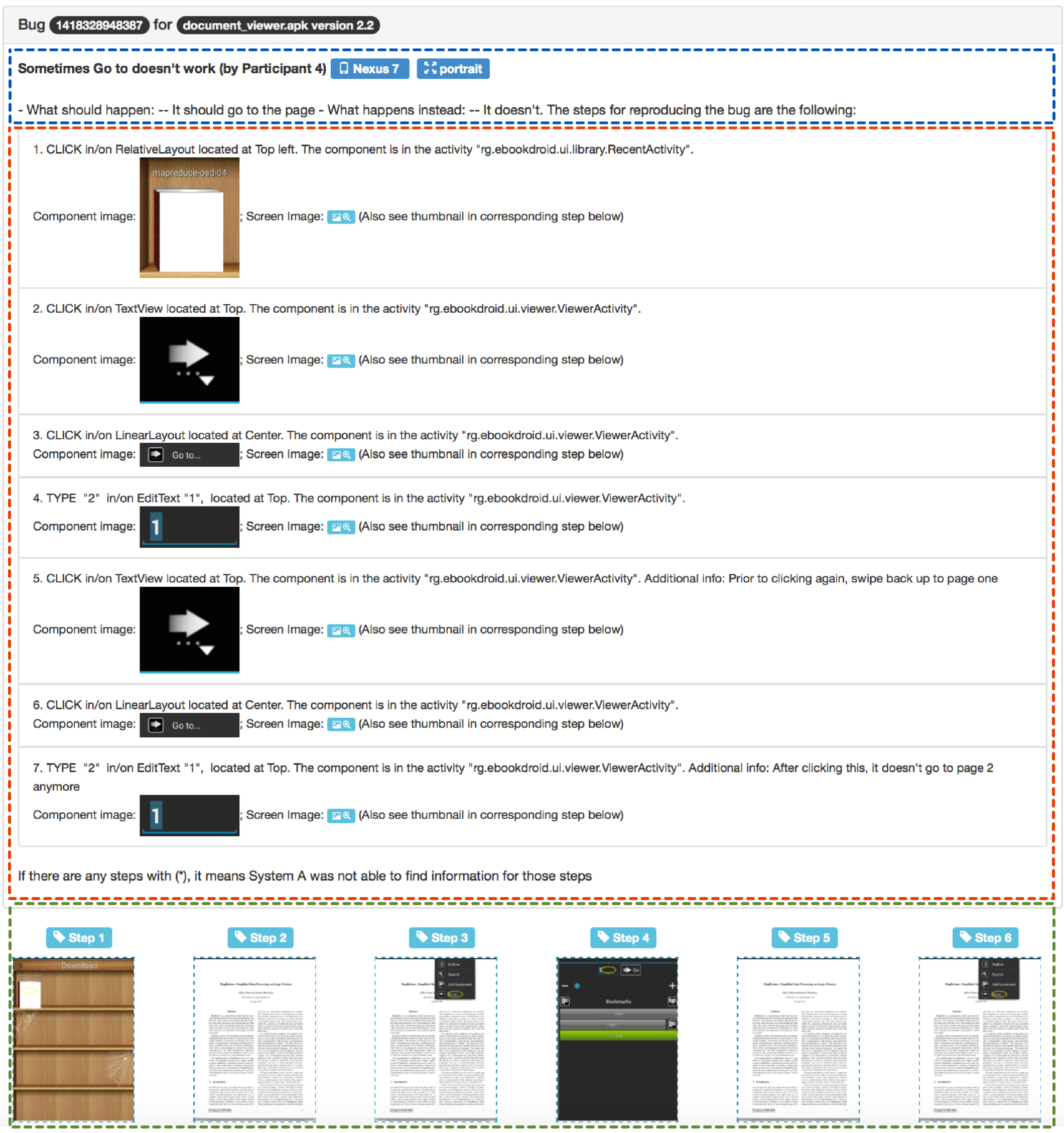}
\caption{Example FUSION Bug Report}
\label{report}
\end{figure}

      The \emph{Auto Completion Engine} saves each step to the database as reporters complete bug reports.  Once a reporter finishes filling out the steps and completes the data entry process, a screen containing the final report, with an automatically assigned unique ID, is presented to the reporter and saved to the database for a developer to view later (see Figure \ref{report} for an example report from Document Viewer). 

       The Report presents information to developers in three major sections: First, preliminary information including the report title, device, and short description (shown in Figure \ref{report} in blue).  Second, a list of the Steps with the following information regarding each step is displayed (highlighted in blue in Figure \ref{report}): (i) The action for each step, (ii) the type of a component, (iii) the relative location of the component, (iv) the activity Java class where the component is instantiated in the source code, and (v) the component specific screenshot.  Third, a list of full screen-shots corresponding to each step is presented at the bottom of the page so the developer can trace the steps through each application screen (this section is highlighted in green in Figure \ref{report}).

\section{Design of the Experiments}
\label{sec:study}
The two major design goals behind FUSION are: 1) to facilitate and encourage reporters to submit useful bug reports for Android applications; 2) to provide developers with more actionable information regarding the bugs contained within these reports. In order to measure how effective FUSION is at achieving these goals, we have evaluated two major aspects of our  approach: 1) \textit{the quality of the bug reports produced by the system}, and 2) \textit{the user experience of reporters and developers using FUSION}. To this end, we investigated the following research questions (RQs):
  
  \begin{itemize}
\item \textbf{RQ$_1$}: \textit{What information fields in bug reports are useful when reporting bugs in Android apps?}
\item \textbf{RQ$_2$}: \textit{Is FUSION easier to use for reporting bugs than traditional bug tracking systems?}
\item \textbf{RQ$_3$}: \textit{Are FUSION reports easier to use for reproducing bugs than traditional bug reports?}
\item \textbf{RQ$_4$}: \textit{Do bug reports generated with FUSION allow for faster bug reproduction compared to reports submitted using traditional bug tracking systems?}
\item \textbf{RQ$_5$}: \textit{Do developers using FUSION reproduce more bugs compared to traditional bug tracking systems?} 

\end{itemize}

The five RQs were investigated with empirical studies representing two maintenance activities involving reporting and reproduction of real bugs in open source apps. In the following subsections we will describe the context of the two studies (i.e., Android apps and bug reports) and the details of each study.

\subsection{ Context: Bug Reports Used in the Studies}
	In order to properly evaluate FUSION when reporting and reproducing bug reports from real world bugs, we manually selected bug reports from Android Open Source apps at F-Droid \cite{fdroid}. We crawled the links of the issue tracking systems of the apps and then manually inspected the bug reports for each project where F-droid had a linked issue tracker.  The criteria for selecting the bug reports were the following: 1) bugs that are reproducible given the technical constraints of our FUSION implementation; 2) bugs of varying complexity, requiring at least three steps of user interaction in order to be manifested; and 3) bugs that are reproducible on the Nexus 7 tablets utilized for the user study. Details of these bug reports can be found in Table \ref{tab:bug-reports} and links can be found in our replication package outlined in Section \ref{sec:rep-pack} and available online at \cite{appendix}.  
	
	FUSION targets bug reports that can be described in terms of GUI events and are not context dependent.  For instance, some bugs are triggered when changing the orientation of the device, or are context dependent (i.e., the bug depends on the network signal quality, GPS location, etc.). We do not claim that the FUSION approach works for all types of Android bugs, but rather acknowledge and give examples of the current limitations in Section \ref{sec:limitations}.

\subsection{Evaluating User Experience, Preferences and Programming Background}

 For both studies, in addition to collecting time information for the creation and reproduction of the bug reports, we collected responses to a set of questions, outlined in Table \ref{tab:ux-questions}.  The questions focused on three different aspects: 1) user preferences, 2) user experience and 3) demographic background, including programming experience.  The user preference related questions (UP questions in Table \ref{tab:ux-questions}) were formulated based on the user experience honeycomb originally developed by Peter Morville \cite{Morville:04} and posed to participants as free-form text entry fields.   The usability was evaluated by using statements based on the SUS usability scale by John Brooke \cite{Brooke:96}. These statements are labeled in Table \ref{tab:ux-questions} with UX. Programming experience was scored by the participant on an extended Likert scale (1 representing a strong disagreement and 10 representing strong agreement). The background information questions are based on the programming experience questionnaire developed by Feigenspan et al \cite{Feigenspan:ICPC12}. For the analysis of the free-form questions, one of the authors analyzed and categorized the answers manually.  Due to space limitations, Table \ref{tab:ux-questions} presents a subset of the questions posed to study participants.  The full set of questions can be found online in the replication package for this work \cite{appendix}.

\subsection{Study 1: Reporting Bugs with FUSION}

\begin{table}[t]
\caption{Questions and statements for evaluating the user experience of the bug tracking systems and the bug reports generated with the analyzed systems.}
\vspace{-0.3cm}
\scriptsize
\begin{center}
\begin{tabular}{|c|p{7cm}|}
\hline
\textbf{Id}&\textbf{Question} \\ \hline
UP1&What information from this \texttt{<system>} did you find useful for reporting/reproducing the bug?  \\ \hline 
UP2&What other information (if any) would you like to see in this \texttt{<system>}?  \\ \hline 
UP3&What elements do you like the most from this \texttt{<system>}?  \\ \hline 
UP4&What elements do you like the least from this \texttt{<system>}?  \\ \hline 
UX1& I think that I would like to have this type of bug report/system frequently.\\ \hline
UX2&I found this type of bug report/system unnecessarily complex.\\ \hline
UX3&I thought this type of bug report/system was easy to read/use.\\ \hline
UX4&I found this type of bug report/system very cumbersome to read/use.\\ \hline
UX5&I thought the bug report/system was really useful for reporting/reproducing the bug.\\ \hline 
\end{tabular}
\end{center}
\label{tab:ux-questions}
\vspace{-0.3cm}
\end{table}%

	 The \textit{goal} of the first study is to assess whether FUSION's features are useful when reporting bugs for Android apps, which aims to answer \textbf{RQ$_1$} \& \textbf{RQ$_2$}. In particular, we want to identify whether the auto-completion steps and in-situ screenshot features are useful when reporting bugs. To accomplish this, we recruited eight students (four undergraduate or \textit{non-experts} and four graduate or \textit{experts}) at the College of William and Mary to construct bug reports using FUSION and Google Code Issue Tracker (GCIT) --- as a representative of traditional bug tracking systems--- for the real world bugs from the reports shown in Table \ref{tab:bug-reports}. The four graduate participants had extensive programming backgrounds. Four participants constructed a bug report for each of the 15 bugs in Table \ref{tab:bug-reports} using FUSION prototype, and four participants reported bugs using the Google Code Issue Tracker Interface.  The participants were distributed to the systems such that two non-experts and two programmers evaluated both systems. In total the participants constructed 120 bug reports, 60 using FUSION and 60 using GCIT.  Participants used a Nexus 7 tablet with Android 4.4.3 KitKat installed to reproduce the bugs.
	 
	 One challenge in conducting this first study is illustrating the bug to the participants without introducing bias from the original bug report.  To accomplish this, we created short videos of the steps to reproduce every bug using the fewest number of actions possible.  After viewing the video each participant was asked to confirm their knowledge of the bug by reproducing it on a Nexus 7 tablet, with a study proctor confirming the reproduction.  Then the participants filled out a bug report for each of the 15 bugs for the system to which they were assigned.  During the report collection, the names of the bug reporting systems were anonymized to ``System A" for FUSION and ``System B" for GCIT. The users were provided with a short tutorial regarding how to enter bugs for each system so as not to introduce bias towards any reporting system.   After the creation of the bug reports, users were asked to answer the questions listed in Table \ref{tab:ux-questions} in an online survey.  Results of this study and the corresponding \textbf{RQ$_1$} \& \textbf{RQ$_2$} are presented in Section \ref{sec:res-study1}.

  \begin{table*}[tb]
\centering
\scriptsize
\caption{Summary of the bug reports used for the empirical studies.  GDE = Gui Display Error, C = Crash, DIC = Data Input/Calculation Error, NE = Navigation Error.}
\label{tab:bug-reports}
\setlength{\tabcolsep}{0.3em}
\begin{tabular}{ | l | p{20pt} | p{300pt} | p{30pt} | p{40pt} | }
\hline
\textbf{App} & \textbf{Bug ID} & \textbf{Description} & \textbf{Min \#steps} & \textbf{Bug Type} \\ \hline
A Time Tracker & 24 & Dialog box is displayed three times in error. & 3 & GDE \\ \hline
Aarddict & 106 & Scroll Position of previous pages is incorrect. & 4-5
 & GDE\\ \hline
ACV  & 11 & App Crashes when long pressing on sdcard folder. & 5 & C \\ \hline
Car report & 43 & Wrong information is displayed if two of the same values are entered subsequently & 10 & DIC\\ \hline
Document Viewer & 48 & Go To Page \# number requires two entries before it works & 4 & NE \\ \hline
DroidWeight & 38 & Weight graph has incorrectly displayed digits & 7 & GDE \\ \hline
Eshotroid & 2 & Bus time page never loads. & 10 & GDE/NE  \\ \hline
GnuCash & 256 & Selecting from autocomplete suggestion doesn't allow modification of value & 10 & DIC \\ \hline
GnuCash & 247 & Cannot change a previously entered withdrawal to a deposit. & 10 & DIC \\ \hline 
Mileage & 31 & Comment Not Displayed. & 5 & GDE/DIC \\ \hline 
NetMBuddy & 3 & Some YouTube videos do not play. & 4 & GDE/NE \\ \hline 
Notepad & 23 & Crash on trying to send note. & 6 & C  \\ \hline 
OI Notepad & 187 & Encrypted notes are sorted in random when they should be ordered alphabetically & 10 & GDE/DIC \\ \hline 
Olam & 2  & App Crashes when searching for word with apostrophe or just a "space" character & 3 & C \\ \hline 
QuickDic & 85 & Enter key does not hide keyboard & 5 & GDE \\ \hline
\end{tabular}
\vspace{-0.5cm}
\end{table*} 
   
 \subsection {Study 2: Reproducibility of Bug Reports}
 
 The \emph{goal} of Study 2 is to evaluate the usability and preferences of developers using FUSION to report bugs as well as the ability of our proposed approach to improve the reproducibility of bug reports, thus answering \textbf{RQ$_3$}-\textbf{RQ$_5$}. In particular, we evaluated the following aspects in FUSION and traditional issue trackers: 1) usability when using the bug tracking systems' GUIs for reading bug reports, 2) time required to reproduce reals bugs by using the bug reports, and 3) number of bugs that were successfully reproduced.  Both the reports generated during Study 1, using FUSION and GCIT, and the original bug reports (Table~\ref{tab:bug-reports}) were evaluated by a new set of 20 participants through attempted bug reproduction on physical devices.
 
The participants were graduate students from the Computer Science Department at College of William and Mary, all of whom are familiar with the Android platform and are experienced programmers. All participants were compensated \$15 USD for their efforts.  Each user evaluated 15 bug reports, six from FUSION, six from GCIT, and three original.  135 reports were evaluated (120 from Study 1 plus the 15 original bug reports) and were distributed to the 20 participants in such a way that each bug report was evaluated by two different participants (the full design matrix can be found in our replication package \cite{appendix}).  Each participant evaluated only one version of a bug report for a given bug, because the learning effect dictates that after a user reproduces a bug once, they will be capable of reproducing it easily in subsequent attempts with other bug reports. 
	
During the study, the participants were sent links corresponding to the reports for which they were tasked with reproducing the bug.  Each participant was loaned a Nexus 7 tablet with Android 4.4.3 KitKat installed; the apps were preinstalled in the devices.  For each bug report, the users attempted to recreate the bug on the tablet using only the information contained within the report.  The users timed themselves in the reproduction for each bug, with a ten minute time limit.  If a participant was not able to reproduce a bug after ten minutes, that bug was marked as \textit{not-reproduced}.  A proctor monitored the study to judge whether participants successfully reproduced a given bug.  After the users attempted to reproduce all 15 bugs assigned to them, they were asked to fill out an anonymous online survey for each type of the bug report they utilized, containing the UX and UP questions listed in Tab \ref{tab:ux-questions}.
For the analysis, we used descriptive statistics to analyze the responses for the UX statements, the time for reproducing the bugs, and the number of successful reproductions. 
Results for \textbf{RQ$_2$}-\textbf{RQ$_4$} are presented in Section \ref{sec:res-study2}.

\section{Results and Discussion}
\label{sec:results}
	In this section we report the results for both studies conducted in our evaluation and outline the major findings.  For a complete dataset and overview of results, including all statistics and user responses, please see our replication package in Section \ref{sec:rep-pack} and online at \cite{appendix}. 

\subsection{Bug Reporting Results from Study 1}
\label{sec:res-study1}
	
	First, we present quantitative and qualitative information based on the time taken to create bug reports and responses from participants in Study 1 in order to answer \textbf{RQ$_1$} \& \textbf{RQ$_2$}.  In regard to the general usefulness of FUSION as a reporting tool, there are two clear trends that emerge from the user responses: 1) \textit{Reporters generally feel that the opportunity to enter extra information in the form of detailed reproduction steps helps them more effectively report bugs}; 2) \textit{Experienced reporters tended to appreciate the value and added effort of adding extra information compared to inexperienced reporters.}  These trends echo the bug creation time results, and there are several statements made by participants that confirm these claims.   For instance, one response from an experienced user to UP1 was the following: ``The GUI component form and the action/event form have been very useful to effectively report the steps."; however, a response to the same question by an inexperienced reporter was, ``I liked the parts where you just type in the information."  One encouraging result during Study 1 is that FUSION was able to auto-suggest all of the reproduction steps without gaps (i.e., auto-completion did not miss any steps) in 11 of 60 bug reports generated.  This means that using the information for the steps contained with FUSION database, extracted during the dynamic execution of an app, \textit{FUSION was able correctly suggest all of the steps to the participant creating a report} and a replayable script can be generated. This would not be possible for GCIT or any other bug tracking system. In summary we can answer \textbf{RQ$_1$} as follows: \textbf{While reporters generally felt that the opportunity to enter extra information using FUSION increased the quality of their reports, inexperienced users would have preferred a simpler web UI.} 
	
	With regard to the time statistics reported in Table \ref{tab:creation-res}, it generally took experienced reporters a similar amount of time to create reports for both systems.  However, inexperienced reporters reported bugs much more quickly with GCIT compared to FUSION.  These results are not surprising, as experienced reporters understand the importance of providing detailed information in bug reports and thus are more likely to create detailed natural language bug reports using \textit{both} GCIT and FUSION.  On the other hand, the results show inexperienced reporters are more likely to create superficial reports using GCIT.  While it did take inexperienced reporters a longer amount of time to create FUSION reports, the creation times were still reasonable and doesn't necessarily reflect poorly on the system.  In fact, these results suggest that FUSION forced even inexperienced reporters to create more detailed, reproducible bug reports, and this is confirmed in the reproduction results.  Thus, we can answer \textbf{RQ$_2$} as follows: \textbf{FUSION was about as easy to use as the GCIT for experienced participants but was more difficult for inexperienced participants to use compared to GCIT.}

\subsection{Bug Reproduction Results from Study 2}
\label{sec:res-study2}
\begin{figure}[t]
\begin{center}
\includegraphics[width=\linewidth]{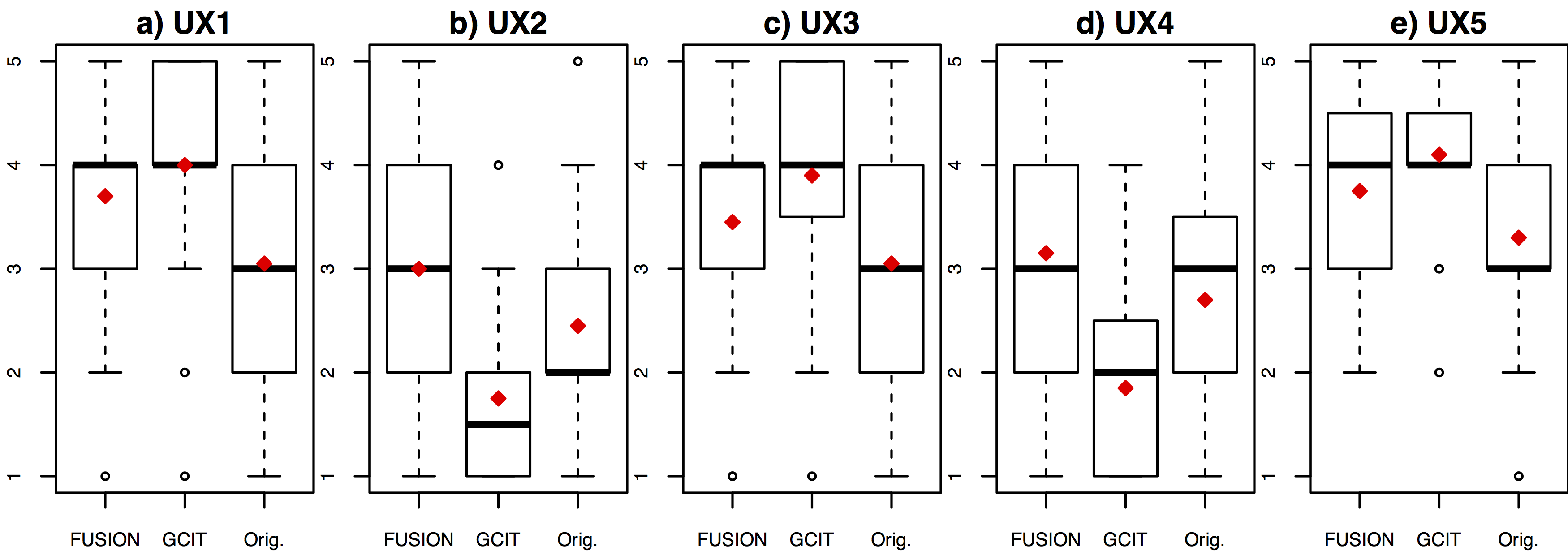}
\vspace{-0.8cm}
\caption{Answers to the UX-related questions in RQ$_3$}
\label{fig:rq2-ux}
\end{center}
\vspace{-0.5cm}
\end{figure}

\begin{figure}[t]
\begin{center}
\includegraphics[width=\linewidth]{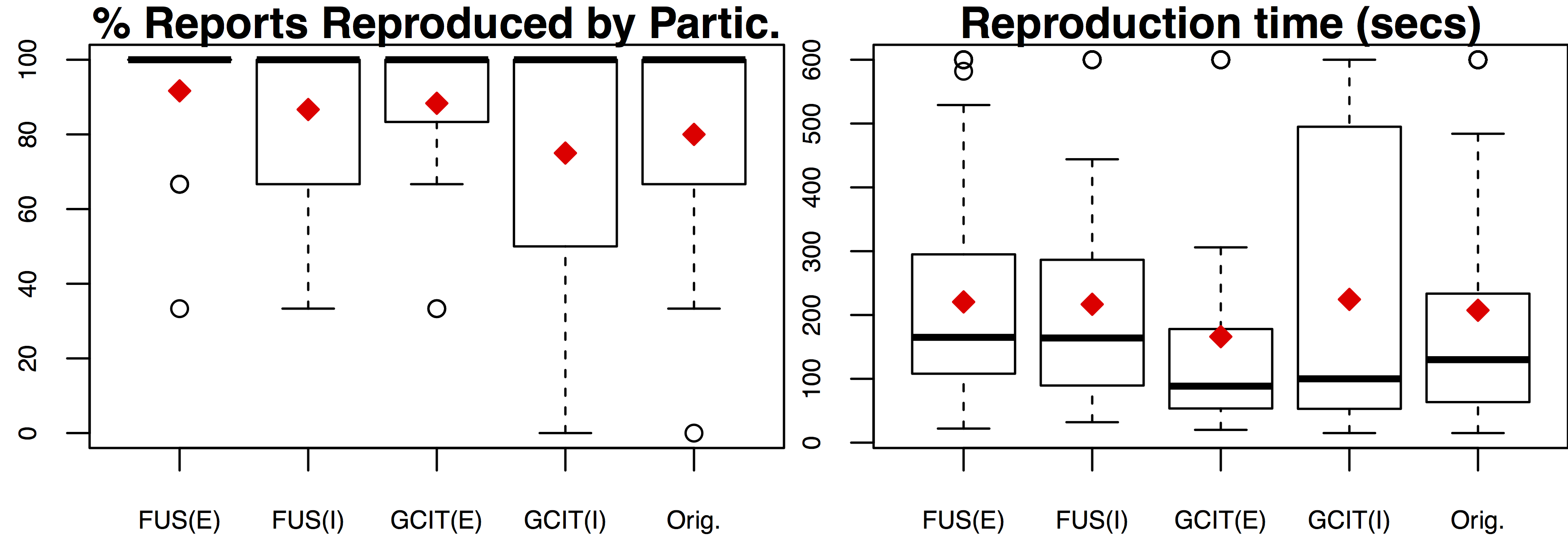}
\vspace{-0.8cm}
\caption{Percentage of bug reports reproduced by each participant (left) for RQ$_3$, and individual bug reproduction time (right) for RQ$_4$.}
\label{fig:rq2-rep_time}
\end{center}
\vspace{-0.3cm}
\end{figure}

\begin{table}[t]
\scriptsize
\small
\caption{Average Bug Report Creation Time: (EX) = Experienced Participant, (IEX) = Inexperienced Participant, Times are reported in (mm:ss) format.}
\label{tab:creation-res}
\begin{tabular}{| p{33pt}  | p{37pt}  | p{37pt}  | p{37pt}  | p{37pt}  |} \hline
                              & Participant \#1 (EX) & Participant \#2 (EX) & Participant \#3 (IEX) & Participant \#4 (IEX) \\ \hline
FUSION: & 5:14                          & 5:20                          & 10:40                           & 4:59                            \\ \hline
                              & Participant \#5 (EX) & Participant \#6 (EX) & Participant \#7 (IEX) & Participant \#8 (IEX) \\ \hline
GCIT:    & 3:17                          & 6:39                          & 1:14                            & 1:46                  \\         
\hline
\end{tabular}
\end{table}

The boxplots in Figures \ref{fig:rq2-ux} and \ref{fig:rq2-rep_time} summarize the results for \textit{Study 2}. In particular, the figures depict the answers to the bug report usability statements (Figure \ref{fig:rq2-ux}), percentage of bug reports reproduced successfully by the participants (Figure \ref{fig:rq2-rep_time}-left), and time required to reproduce the bug reports (Figure \ref{fig:rq2-rep_time}-right).  In the case of reproduction time, because some of the reports were not reproduced during a 10 minute time slot, we set the reproduction time to 600 seconds for visualization and analysis purposes.

  The usability scores in Figure \ref{fig:rq2-ux} show that most users agree that they would like to use FUSION's bug reports frequently, however, several users also found the bug reports to be unnecessarily complex, and some users found the bug reports difficult to read/comprehend.  Most users agreed that they thought FUSION bug reports were useful for helping to reproduce the bugs.  GCIT had the best usability scores out of the three systems, whereas the Original bug reports had the lowest usability scores.  According to user preference feedback for UP3, we received encouraging responses; for instance: ``The detail steps to find where to find the next steps was really useful and speeded up things."; ``The images of icons help a lot, especially when you have a hard time locating the icons on your screen."  However, users also expressed issues with the FUSION report layout: ``Sometimes the steps were too overly specific/detailed."; ``The information, while thorough, was not always clear"; ``If there are steps missing, it is confusing because it is otherwise so detailed."  
  
Based on these responses we can answer \textbf{RQ$_3$} as follows: \textbf{According to usability scores, participants generally preferred FUSION over the original bug reports when reproducing bugs, but generally preferred GCIT to FUSION by a small margin.  The biggest reporter complaint regarding FUSION was the organization of information in the report.}

Figure \ref{fig:rq2-rep_time} details reproducibility results for bug reports written with FUSION by experienced  (i.e., FUSE(E)) and non-experienced participants (i.e., FUS(I)), reports written in GCIT by experienced (i.e., GCIT(E)) and non-experienced participants (i.e., GCIT(I)), and original reports (i.e., Orig). According to Figure \ref{fig:rq2-rep_time}, the average time to reproduce for the two flavors of FUSION were 220.5 and 216.8 seconds respectively for FUS(E) and FUS(I).  Surprisingly, the FUS(I) reports had a smaller average reproduction time than the FUS(E) reports. GCIT reports (E) \& (I) had an average time to reproduce of 166.07 and 224.45 seconds respectively. While this result shows that participants took longer to reproduce FUSION reports, this is to be expected as they had to read and process the extra information regarding the reproduction steps. However, reproduction time of inexperienced reporters with  FUSION is lower than GCIT. While there is no strong correlation as to which system is more capable of creating reproducible reports for complex bugs, we do see that the complex bugs generally have more instances where they are not reproducible, which is to be expected. 

Based on these results we can answer \textbf{RQ$_4$} as follows: \textbf{Bug reports generated with FUSION do not allow for faster reproduction of bugs compared bug reports generated using traditional bug tracking systems such as the GCIT} .

In terms of reproducibility, overall, the reports generated using FUSION were more reproducible than the reports generated using GCIT with only 13 of the 120 bug reports from FUSION being non-reproducible compared to 23 of the 120 reports from GCIT being non-reproducible. The bug report type with the lowest number of non-reproducible cases is FUS(E), whereas the report type with the highest number of non-reproducible cases is GCIT(I).  One encouraging result is that when inexperienced participants created bug reports in \textit{Study 1}, participants in \textit{Study 2} seemed to have a much easier time reproducing the reports from FUSION (I), which only had 8 non-reproducible cases, compared to GCIT(I) which had nearly twice as many, 15, non-reproducible cases. This means that for reporters classified as inexperienced FUSION could greatly improve the bug report quality.  Both of the individual FUSION bug report types (I) and (E) had a lower number of non-reproducible cases than the Original bug reports as well.  However, a direct comparison cannot be made here, as each original bug report was tested (attempted reproduction) four times, compared to two times for FUSION and GCIT bug reports.

Therefore, based on these results we can answer \textbf{RQ$_5$} as follows: \textbf{Developers using FUSION are able to reproduce more bugs compared to traditional bug tracking systems such as the GCIT.}

\subsection{Lessons Learned}

	The major lessons that can be gleaned from the results of \textit{Study 1}, which should be taken into account in future research and issue tracker design, are \textbf{1)} \textit{Intuitive UI design is extremely important to enhance the usability of issue trackers for reporters}, and \textbf{2)} \textit{presenting users with a structured reporting mechanism, such as that in FUSION, can increase the quality of bug reports, even for inexperienced participants}.  If an issue tracker is able to successfully combine features that address both of these lessons, the result will be a system that places less burden on the reporter and produces more useful bug reports.
		
	There are two major lessons that emerge from the results of \textit{Study 2}: \textbf{1)} \textit{the design of the report should be specifically suited to the maintenance task required}.  Several participants complained of overly specific or detailed information during the second study, and this information may have been more suited to a fault-location task.  In our study we focused on reproduction to gauge bug report quality as it is well known that if a developer can reproduce a bug there is a much higher chance that they will be able to fix and patch it \cite{4Joorabchi:MSR14, 3Bettenburg:FSE08, 15Breu:CSCW10}.  However, based on the user experience and preference results from \textit{Study 2}, it may be beneficial to present information to developers in stages (e.g., first present reproduction steps, then more detailed code-information for fault location).  Lesson \textbf{2)} \textit{there is a clear-trade off between time and bug reproduction ability in more detailed bug reports such as those produced by FUSION}.  FUSION reports were generally more accurate, but took slightly longer to reproduce; however, this is a tradeoff developers would be willing to make in the competitive mobile app marketplace.

\subsection{Limitations}
\label{sec:limitations}
Currently, the DFS implementation in FUSION only supports the \textit{click/tap} action.  Another option to gather runtime program information would be to record app scenarios and replay them while collecting program data or using language modeling based approaches for scenario generation \cite{Linares:MSR15} .  However, we forwent such an approach in favor of the fully automatic DFS application exploration and constructing a completely off-device issue tracking system that may be able to describe bugs a record and replay approach might miss.  Part of our immediate plan for future work includes adding support for more gestures to our DFS engine.  FUSION is currently not capable of capturing certain contextual app  information such as a change in device orientation or network state.  However, this can be mitigated by the fact that reporters can enter such contextual information in the free-form text field associated with each step.  FUSION is also limited in the types of bugs that it can report, currently supporting functional bugs that can be uncovered using only GUI-Gestures such as tap, long-touch, swipe and type.  It is important to note that even though the systematic section engine is not able to perform and capture gestures other than tap, these gestures can still be reported using FUSION.

\section{Threats to Validity}
\label{sec:threats}
	Threats to internal validity concern issues with the legitimacy of causal relationships inferred.  In the context of our studies, threats come from potentially confounding effects of participants.   First, we assume that undergraduate students without a CS background but who have experience using Android devices are representative of \textit{non-expert} testers.  We believe this is a reasonable assumption given the context, as most \textit{non-expert} testers will only have a ``working" knowledge of the app and platform.  We also assumed graduate students with Android experience were reasonable substitutes for developers.  Again, we believe this is reasonable given that all four of the ``experienced" participants in \textit{Study 1} indicated they had extensive programming backgrounds and reasonable Android programming experience (at least 4 on the scale where 10 represents ``Very experienced").  Likewise, the participants in \textit{Study 2} indicated that they all had extensive programming backgrounds, and 13 of the 20 participants had reasonable Android programming experience.

	Threats to external validity concern the generalizability of the results.  The first threat of this type relates to the bug reports and Android apps used in our study.  We evaluated FUSION on only 15 bug reports from 14 different applications from the F-droid \cite{fdroid} marketplace.  In order to increase the generalizability of the results we aimed at selecting bug reports of varying type and complexity from apps representing different categories and functions.  During our study we utilized only one device type, a Nexus 7 tablet, in order to standardize results across participants.  However, there is nothing limiting FUSION from being utilized on several different Android devices from varied manufacturers.  We concede that FUSION is not suited for reporting all types of bugs (e.g., nuanced performance bugs, context dependent bugs), however, we conjecture that any type of bug that can be reported with a traditional issue tracking system can be reported with FUSION.

\section{Conclusion and Future Work}
\label{sec:concl}
Prior research highlights an inherent lexical gap that exists between reporters of bugs and developers.  To help overcome this, we introduced FUSION, a novel bug reporting approach, that takes advantage of program analysis techniques and the event-driven nature of Android applications, in order to help auto-complete the reproduction steps for bugs.  Results from our comprehensive evaluation show FUSION is able to produce more reproducible bug reports than traditional issue tracking systems.  We hope our work on FUSION encourages a new direction of research aimed at \textit{improving reporting systems}.  In future work, we aim to improve our DFS engine through supporting more gestures, to explore adding more specific program information in reports for quicker/automatic fault localization, and to use FUSION as a tool for reporting feature requests.

\section{Acknowledgments}
\label{sec:ack}
This work is supported in part by the NSF CCF-1218129 and NSF CCF-1253837 grants.  Any opinions, findings, and conclusions expressed herein are the authors' and do not necessarily reflect those of the sponsors.   We would like to thank Martin White for his invaluable guidance at the outset of this project and the anonymous reviewers for their insightful comments which greatly improved this paper.

\section{The FUSION Replication Package}
\label{sec:rep-pack}
In order to enhance the reproducibility of the results obtained in our evaluation of FUSION, we offer a replication package containing a live instance of FUSION running on the web, and the full dataset of all results obtained during our comprehensive empirical evaluation.  The FUSION replication package has been successfully evaluated by the Replication Packages Evaluation Committee and found to meet expectations.  The replication package is accessible at \cite{appendix} and we outline its contents and utility in this section.

\subsection{Contents of the Replication Package}

All of the replication materials for this work can be accessed through the project webpage \cite{appendix}, this website contains the following materials:

\begin{itemize}

\item{\textbf{Project Overview:}  This section of the webpage contains a high-level overview of the FUSION project as well as author information and a brief description of how to navigate the site.}
\item{\textbf{Component I: FUSION} This section of the webpage describes the FUSION tool in detail, including information regarding the tools used in our implementation of the various components.  This section also contains links to live instances of the FUSION reporting system \cite{fusion-reporting_system} and report viewer \cite{fusion-report_viewer} accessible through the web.  This section also contains a video demonstration of FUSION in action, and documentation regarding how to use the interface for creating and viewing reports.}
\item{\textbf{Component II: Results \& Reproduction:}  This section contains a detailed discussion of the results obtained from our empirical evaluation of FUSION, including figures and statistics not reported in this paper due to space constraints.  This section also offers links to download the complete dataset from our studies in both \texttt{.xlsx} and \texttt{.csv} format.}

\end{itemize}

\subsection{Component I: Understanding and Using FUSION}

\subsubsection{Tools Used for FUSION's Implementation}

In this component, we provide the tools used for our implementation of FUSION along with links to the tools themselves, which we outline below:

\textbf{Tools Used To Implement the Static Analyzer (Primer):}

\begin{itemize}
\item{\textbf{APKTool:} a tool for reverse engineering Android apk files.}
\item{\textbf{Dex2jar:} A conversion tool for .dex files and .class files.}
\item{\textbf{jd-cmd:} A command line Java Decompiler.}
\end{itemize}

\textbf{Tools Used To Implement the Dynamic Program Analyzer (Engine):}

\begin{itemize}
\item{\textbf{Android Debug Bridge (adb):} A universal tool for communicating with Android devices and emulators.}
\item{\textbf{Hierarchy Viewer:} A tool for examining and optimizing Android user interfaces.}
\item{\textbf{UIAutomator:} A tool that provides a set of APIs to build UI tests for Android applications and aid in interacting with GUI Components.}
\end{itemize}

\textbf{Tools Used To Implement the FUSION Web Interface:}

\begin{itemize}
\item{\textbf{Bootstrap:} HTML, CSS, and JavaScript framework for developing web applications.}
\item{\textbf{MySQL:} relational database.}
\end{itemize}

\subsubsection{Live Web Instance of FUSION}

	In order to promote the reproducibility of the results obtained for our empirical study, we provide a live instance of both the FUSION reporting system and the report viewer running on the web.  These instances contain the 14 open source applications and all of the bug reports created and evaluated by participants during the empirical studies.   At the time of publication, we do not offer access to the static and dynamic analysis components of FUSION due to ongoing development associated with future research projects, however, we may package and release these as closed-source tools at the request of researchers who need access for purpose of comparison, and the authors will do their best to answer any questions regarding the implementation of these tools if contacted.  This section in the replication package contains full documentation with screenshots, as well as a video demonstration outlining how to use FUSION.
	
\subsection{Component II: FUSION Results and\\ Dataset}

	In order to promote transparency for this work and facilitate researchers working on similar projects, we provide the full dataset collected during the empirical studies conducted to evaluate FUSION.  This dataset contains all of the time, reproduction, user experience, and user preference results from our study.  For convenience, we offer the results in either \texttt{.xlsx} or \texttt{.csv} format.  The excel workbook is broken into sheets each with different results outlined on each sheet, and the \texttt{.csv}  representation is broken into separate files each containing different results.  The list of results contained in these sheets/files is as follows: 

\begin{itemize}
\item{\textbf{Study 1:} User Experience (Likert Scale Results)} 
\item{\textbf{Study 1:} User Preference (Open Responses)} 
\item{\textbf{Study 1:} Bugs Created (Full List of FUSION and GCIT Report Numbers \& Links)}
\item{\textbf{Study 1:} Bug Creation Time Results (Time Statistics)}
\item{\textbf{Study 1:} Participant Programming Experience (Likert Scale Results)}
\item{\textbf{Study 2:} User Experience (Likert Scale Results)}
\item{\textbf{Study 2:} User Preference (Open Responses)}
\item{\textbf{Study 2:} Bug Reproduction (Full Time/Reproduction Results)}
\item{\textbf{Study 2:} Aggregated Bug Reproduction Results (Summarized Time/Reproduction Results)}
\item{\textbf{Study 2:} Participant Programming Experience (Likert Scale Responses)}
\end{itemize}

\balance

\balancecolumns

\end{document}